\documentclass[11pt]{article}
\usepackage[colorlinks=true, linkcolor=black!50!blue, urlcolor=blue, citecolor=blue, anchorcolor=blue]{hyperref}
\usepackage[font=small,labelfont=bf,margin=0mm,labelsep=period,tableposition=top]{caption}
\usepackage[a4paper,top=3cm,bottom=2.5cm,left=2.5cm,right=2.5cm,bindingoffset=0mm]{geometry}

\usepackage{graphicx,placeins}
\usepackage{float}
\usepackage{afterpage}
\usepackage{epsfig,cite}
\usepackage{amssymb}
\usepackage{amsmath}
\usepackage{dsfont}
\usepackage{multirow}
\usepackage{url}
\usepackage{xcolor,colortbl}
\usepackage{float}
\usepackage{afterpage}
\usepackage{url}
\usepackage{hyperref}
\usepackage{booktabs}
\usepackage{mathrsfs}

\usepackage{tikz}
\usepackage{tikz-3dplot}
\usepackage[compat=1.0.0]{tikz-feynman}

\usepackage{enumitem}
\usepackage{hyperref}
\usepackage{cite}

\usepackage{pifont}

\usetikzlibrary{shapes, arrows}
\usetikzlibrary{decorations.pathreplacing}
\usetikzlibrary{positioning, calc}
\tikzstyle{fitted} = [rectangle, minimum width=5cm, minimum height=1cm, text centered, draw=black, fill=red!30]
\tikzstyle{operations} = [rectangle, rounded corners, minimum width=2cm,text centered, draw=black, fill=red!30]
\tikzstyle{roundtext} = [rectangle, rounded corners, minimum width=2cm, minimum height=0.8cm, text centered, draw=black, fill=red!30]
\tikzstyle{n3py} = [rectangle, rounded corners, minimum width=3cm, minimum height=1cm, text centered, draw=black, fill=green!30]
\tikzstyle{myarrow} = [thick,->,>=stealth]
\tikzstyle{line} =[draw, -latex']
\tikzstyle{decision} = [diamond, draw, fill=red!20, text width=7.5em, text centered,  inner sep=0pt, minimum height=2em, aspect=4]
\tikzstyle{cloud} = [draw, ellipse,fill=green!20, minimum height=2em]
\tikzstyle{inout} = [rectangle, draw, fill=green!20, text width=9.5em, text centered, rounded corners, minimum height=2em, minimum width=10em]
\tikzstyle{block}=[rectangle, draw, fill=blue!20, text width=9.5em, 
                   text centered, rounded corners, minimum height=2em, 
                   minimum width=10em]

\definecolor{darkgreen}{rgb}{0.0, 0.5, 0.13}

\newcommand{\be}{\begin{equation}}
\newcommand{\ee}{\end{equation}}
\newcommand{\bea}{\begin{eqnarray}}
\newcommand{\eea}{\end{eqnarray}}
\newcommand{\bi}{\begin{itemize}}
\newcommand{\ei}{\end{itemize}}
\newcommand{\ben}{\begin{enumerate}}
\newcommand{\een}{\end{enumerate}}

\newcommand{\lp}{\left(}
\newcommand{\rp}{\right)}

\def\gsim{\mathrel{\rlap{\lower4pt\hbox{\hskip1pt$\sim$}}
    \raise1pt\hbox{$>$}}}         
\def\lsim{\mathrel{\rlap{\lower4pt\hbox{\hskip1pt$\sim$}}
    \raise1pt\hbox{$<$}}}         

\newcommand{\draft}[1]{}

\usepackage{tabularx}
\newcolumntype{C}[1]{>{\centering\arraybackslash}p{#1}}

\usepackage{amsmath}
\usepackage{amsfonts}
\usepackage{amssymb}
\usepackage{dsfont}
\usepackage{pifont}
\usepackage{booktabs}
\usepackage{graphicx}
\usepackage{epstopdf}
\usepackage{epsfig}
\usepackage{framed}
\usepackage{makeidx}
\usepackage{siunitx}
\usepackage[capitalise]{cleveref}
\usepackage{hyperref}
\usepackage{placeins}
\usepackage[font=small,labelfont=bf]{caption}

\RequirePackage{csquotes}
\usepackage{bbm}

\newcommand\pubdate{\today}

\def\Title#1{\begin{center} {\Large #1 } \end{center}}
\def\Author#1{\begin{center}{ \sc #1} \end{center}}
\def\Address#1{\begin{center}{ \it #1} \end{center}}

\newcommand\pubblock{\rightline{\begin{tabular}{l} \\ 
         \pubdate  \end{tabular}}}
\newenvironment{Abstract}{\begin{quotation}  }{\end{quotation}}
\newenvironment{Presented}{\begin{quotation} \begin{center} 
             PRESENTED AT\end{center}\bigskip 
      \begin{center}\begin{large}}{\end{large}\end{center} \end{quotation}}

\begin{document}
\begin{titlepage}
 \pubblock
\vfill
\Title{DGLAP evolution of parton distributions at approximate N$^3$LO}
\vfill
\Author{Felix Hekhorn$^a$, Giacomo Magni$^{b,c}$}
\Address{ 
    ~${}^a$Tif Lab, Dipartimento di Fisica, Universit\`a di Milano and\\ 
    INFN, Sezione di Milano,
    Via Celoria 16, I-20133 Milano, Italy\\[0.1cm]
    ~${}^b$Department of Physics and Astronomy, Vrije Universiteit, NL-1081 HV Amsterdam\\[0.1cm]
    ~${}^c$Nikhef Theory Group, Science Park 105, 1098 XG Amsterdam, The Netherlands\\[0.1cm]
}
\vfill
\begin{Abstract}
  We present recent progress towards a global determination of
  parton distribution functions (PDFs) at  approximate N$^3$LO (aN$^3$LO) accuracy
  within the NNPDF framework.
  We construct a parametrisation of the $\mathcal{O}(\alpha_s^4)$ QCD splitting
  functions and anomalous dimensions reproducing all known exact results,
  estimate the associated missing and incomplete higher order uncertainties (MHOU and
  IHOU, respectively), and implement it in the open-source DGLAP code {\sc\small EKO}
  enabling PDFs to be evolved at aN$^3$LO accuracy in the NNPDF fitting framework.
  We compare aN$^3$LO calculation of splitting functions with  the results of lower perturbative orders
  and quantify the impact of the various sources of theoretical uncertainties.
\end{Abstract}
\vfill
\begin{Presented}
DIS2023: XXX International Workshop on Deep-Inelastic Scattering and
Related Subjects, \\
Michigan State University, USA, 27-31 March 2023 \\
     \includegraphics[width=9cm]{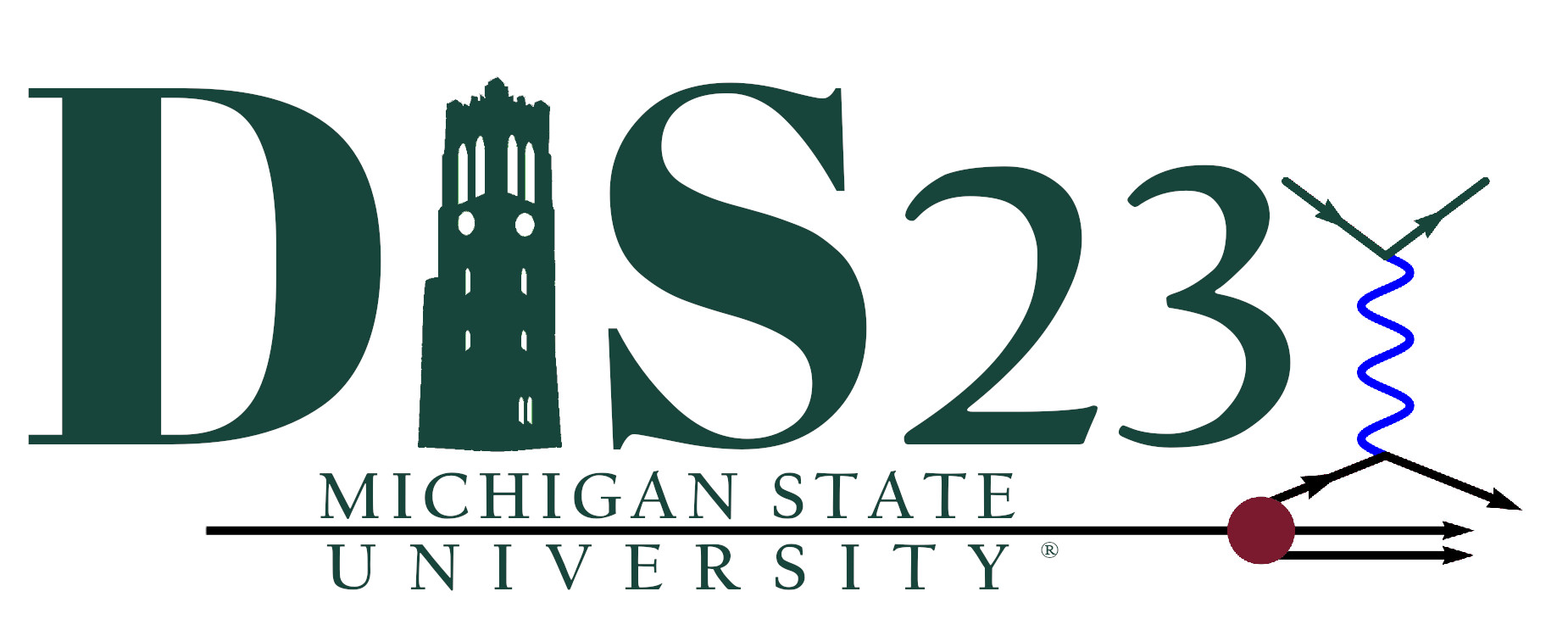}
\end{Presented}
\vfill
\end{titlepage}

\clearpage

\paragraph{Introduction.}
The recent progress in precision measurements at the LHC calls for an equally advanced description from the side of
theoretical calculations~\cite{Amoroso:2022eow}.
In the framework of collinear factorization, theoretical predictions
are separated 
into partonic matrix elements and parton distribution functions
(PDFs)~\cite{Gao:2017yyd}.
In the last years, a significant effort has been made in computing partonic matrix elements 
up to the next-to-next-to-next-to-leading order (N$^3$LO) in the expansion in the strong coupling $\alpha_s$.
However, the perturbative accuracy of the PDFs
in most global determinations is limited to next-to-next-to-leading order (NNLO), which prevents achieving full N$^3$LO for
the hadronic cross-sections~\cite{Caola:2022ayt}.
A first attempt towards PDFs at N$^3$LO based on the MSHT20
fitting framework has been presented in~\cite{McGowan:2022nag}.

Key ingredients in determining PDFs at a given order in the QCD
expansion are the Altarelli-Parisi
splitting functions $P(x, \mu^2)$, currently known exactly up to three loops~\cite{Moch:2004pa,Vogt:2004mw,Blumlein:2021enk}.
While the full N$^3$LO calculation is not available yet, different approximations exist such as the 
large $n_f$ limit~\cite{Davies:2016jie}, the leading color expansion~\cite{Moch:2017uml},
the small-$x$ asymptotic behavior~\cite{Bonvini:2018xvt},
the large-$x$ asymptotic behavior~\cite{Davies:2022ofz},
as well as several Mellin moments~\cite{Moch:2017uml,Moch:2021qrk,Falcioni:2023luc}.
Here we combine all available exact results to construct
an 
approximate compact parametrisation of the N$^3$LO splitting functions
$P_{qq}, P_{gq}, P_{qg},$ and $P_{gg}$.
Unconstrained terms in this parametrization are taken into account
by means of an estimate of the incomplete higher order uncertainties (IHOU).
We also estimate missing higher order uncertainties (MHOU) from scale variations
following the procedure developed in~\cite{AbdulKhalek:2019ihb,AbdulKhalek:2019bux}.
The resulting aN$^3$LO splitting functions are implemented
in the open-source PDF evolution code {\sc\small EKO}~\cite{Candido:2022tld}
and are therefore
integrated in the theory pipeline~\cite{Barontini:2023vmr}
underlying the NNPDF fitting framework~\cite{NNPDF:2021uiq,Ball:2021leu}.

\paragraph{DGLAP evolution at N$^3$LO.}
In Mellin  space one can express the DGLAP
evolution equations in the singlet sector as
\begin{equation}
    \frac{d}{d\ln(\mu^2)} \begin{pmatrix} \Sigma \\ g \end{pmatrix} =
    - \begin{pmatrix} \gamma_{qq} & \gamma_{qg} \\ \gamma_{gq} & \gamma_{gg} \\ \end{pmatrix} 
    \begin{pmatrix} \Sigma \\ g \end{pmatrix}
    \label{eq:dglap_singlet}
\end{equation}
where the anomalous dimensions $\gamma_{ij}(N,a_s(\mu^2))$ are related to the
Altarelli-Parisi splitting functions $P(x)$ by
\begin{equation}
    \gamma(N,a_s(\mu^2)) = - \int\limits_0^1 dx\,x^{N-1}P(x,a_s(\mu^2))
\end{equation}
and which can be expanded in powers of the strong coupling $a_s(\mu^2) = \alpha_s(\mu^2)/4\pi$ by
\be
    \gamma_{ij}(N,a_s(\mu^2)) = a_s \gamma_{ij}^{(0)}(N)
    + a_s^2 \gamma_{ij}^{(1)}(N)
    + a_s^3 \gamma_{ij}^{(2)}(N)
    + a_s^4 \gamma_{ij}^{(3)}(N)
    + \ldots \, .
    \label{eq:ad_expansion}
\ee
Analogous expansions exist for the non-singlet sector evolution equations.
While exact results at NNLO accuracy, $\gamma_{ij}^{(2)}(N)$, have
been known for some time~\cite{Moch:2004pa,Vogt:2004mw,Blumlein:2021enk},
only partial results exist at N$^3$LO.
Note that we defined the usual combination
\be
    \gamma_{qq}(N,a_s(\mu^2)) = \gamma_{NS,+}(N,a_s(\mu^2)) + \gamma_{qq,PS}(N,a_s(\mu^2)) \, ,
    \label{eq:g_qq}
\ee
where $\gamma_{NS,+}$ is the singlet-like non-singlet anomalous dimension,
which arises in the non-singlet sector together with valence-like non-singlet anomalous dimension $\gamma_{NS,-}$
and the valence anomalous dimension $\gamma_{NS,v}$.
The large-$x$ limit of the splitting
functions corresponds in Mellin space to large-$N$,
where the leading contribution for same flavor splitting comes from the harmonics $S_1(N)$.
The small-$x$ region is dictated by the poles at $N=0,1$ depending on the
splitting function
In the following, we describe how the known
partial results at N$^3$LO 
can be used to approximate $\gamma^{(3)}_{ij}$ with
a relatively compact parametrization.

First, we consider the dependency on the number of flavors $n_f$: since this dependency can
be fully predicted, we split the analytical known behavior~\cite{Davies:2016jie}
apart and use it as is.
Second, we address the small-$x$ divergence: here we have to distinguish between the non-singlet 
and the singlet splitting functions, as the former can contain single logarithms
\begin{equation}
    P_{ns}^{(3)}(x) \supset \sum_{k=1}^{6} c^{(k)} \ln^k(x) \, ,
    \label{eq:ns_smallx}
\end{equation}
while the latter features stronger small-$x$ divergences of the type
\be
    P^{(3)}_{ij} \supset \sum_{k=0}^{3} \frac{\ln^{k}(x)}{x} \, ,
    \label{eq:s_smallx}
\ee
which might spoil the perturbative convergence of the $a_s$ expansions
when $x$ tends to 0.
This behavior is well understood and studied in the
context of the BFKL resummation~\cite{Thorne:1999rb,Ball:2005mj,Ball:2017otu}, and at N$^3$LO the first two terms
of the logarithmic expansion are known~\cite{Bonvini:2018xvt}.
In Mellin space these terms correspond to a pole at $N=1$, 
and we remark that known constrains are not sufficient to fix completely
all the divergent terms.
Due to accidental cancellations occurring at NNLO, these logarithms
are larger than the one present in $P^{(2)}_{ij}(x)$ and 
are effectively determining the rise of the splitting functions at small-$x$.
At leading-logarithmic (LL) accuracy, the quark splitting
functions respect the representation symmetry:
\begin{align}
    \gamma_{gq} & \approx \frac{C_F}{C_A} \gamma_{gg} \, , \quad
    &\gamma_{qq,PS} & \approx \frac{C_F}{C_A} \gamma_{qg} \, .
\end{align}
The sub-leading single logarithms of Eq.~(\ref{eq:ns_smallx}) are also present in the 
singlet splitting functions and also known~\cite{Davies:2022ofz},
which allows fixing the leading divergent coefficients of the pole at $N=0$.

Third, we address the large-$x$ behavior where we have to
distinguish the diagonal part and the off-diagonal contributions.
The diagonal terms diverge in $N$-space as 
\cite{Albino:2000cp,Moch:2021qrk}:
\begin{align}
    \gamma_{gg} & \approx A^{(a)}_4 S_1(N) + B^{(a)}_4 + \mathcal{O}(1) \, , \\
    \gamma_{ns}^{(3)} & \approx A^{(f)}_4 S_1(N) - B^{(f)}_4 + C_4 \frac{S_1(N)}{N} - \lp D_4 + \frac{1}{2} A^{(f)}_4 \rp \frac{1}{N} + \mathcal{O}\lp \frac{\ln^k(N)}{N^2}\rp \, ,
\end{align}
where the coefficient $A^{(r)}_4$ is the QCD cusp anomalous dimension 
for the adjoint or fundamental representation.
The coefficient $B^{(r)}_4$ is fixed by the integral of the 4-loop splitting function and has been 
first computed in~\cite{Moch:2017uml} in the large $N_c$ limit, and more recently, 
it has been determined in the full color expansion by computing different N3LO cross sections 
in the soft limit~\cite{Duhr:2022cob}.
For $\gamma_{ns}^{(3)}$ also the coefficients
$C_4,D_4$ are known.
On the other hand, $\gamma_{qq,PS}^{(3)}$ is suppressed at large-$x$ and thus does not constrain 
any divergence~\cite{Soar:2009yh}.
This yields an expansion suppressed by a factor $(1-x)$:
\be
    P_{qq,PS} \approx (1-x)[c_{4} \ln^4(1-x) + c_{3} \ln^3(1-x)] + \mathcal{O}((1-x)\ln^2(1-x)) \, .
\ee
Also the off-diagonal terms $\gamma_{gq}, \gamma_{qg}$ do not contain any 
plus-distributions but may include divergent logarithms~\cite{Soar:2009yh}
of the type of

\begin{equation}
    P_{ij}^{(3)}(x) \supset \sum_{k=1}^{6} c^{(k)} \ln^k(1-x) \, ,
\end{equation}
where the term $k=6$ vanishes.
The values of the coefficient for $k=4,5$
can be estimated from the lower order splitting functions.

The fourth and last known type of constraint comes for the 4 lowest even
Mellin space moments computed
in~\cite{Moch:2021qrk} and including the momentum conservation
which implies
\begin{align}
    \gamma_{qg}(N=2) + \gamma_{gg}(N=2) &= 0 \, , \quad
    &\gamma_{qq}(N=2) + \gamma_{gq}(N=2) &= 0 \, .
\end{align}
Recently, 6 additional moments for $\gamma_{qq,PS}$
were computed~\cite{Falcioni:2023luc}  and are also included in our study. 

To construct the sought-for approximation for $P_{ij}^{(3)}$, we
parametrize 
the difference between the known moments and the above limits as a linear combination of sub-leading functions.
However, due to the limited number of constraints,
we are not able to reconstruct completely the N$^3$LO terms
and an additional source of uncertainties, which we denote
as IHOUs, must be included in our
parametrization.

\paragraph{Theoretical uncertainties on the splitting functions.}
Two different types of theoretical uncertainties can be included in
the splitting function determination.
The first arises from the  missing higher orders in perturbative expansions on $a_s$ in
\cref{eq:ad_expansion} and are denoted by MHOUs.
The second, instead, is due to the incomplete knowledge of 
the N$^3$LO contributions $P^{(3)}_{ij}$ and are denoted
as IHOUs.
We consider the two uncertainties as uncorrelated.

Thanks to the renormalization group invariance, factorization scale variations provide a way to access
MHOUs~\cite{AbdulKhalek:2019ihb,Vogt:2004ns}, as the factorization scale is an unphysical scale.
We can use this mechanism to generate scale-varied anomalous dimensions $\bar \gamma\left(a_s(\rho \mu^2),\rho\right)$
depending on an arbitrary parameter $\rho$ which captures the desired behavior
\begin{equation}
    \bar \gamma\left(a_s(\rho \mu^2),\rho\right) = \gamma\left(a_s(\mu^2)\right) + O\left(\left(\alpha_s(\mu^2)\right)^{5}\right) \, ,
    \label{eq:MHOU}
\end{equation}
and use these variations to evaluate the MHOUs and add them to the theory covariance matrix
of~\cite{AbdulKhalek:2019ihb}.

Concerning IHOUs,
since the available constrains on the singlet N$^3$LO anomalous dimension are not sufficient
to determine their behavior exactly (for instance the poles at $N=0,1$ are not fully known)
we need to account for a possible source of uncertainties arising during the approximation.
This uncertainty can be neglected in the non-singlet case.
We perform a two-step procedure for each different anomalous dimension separately.
First, having accounted for known limits, we solve in Mellin space the linear system associated
to the 4 (10 for $\gamma_{qq,PS}$) known moments, using different functional bases.
Any possible candidate contains 4 (10) elements and is obtained with the following prescription:
\begin{enumerate}
    \item one function, $f_1$, is the leading large-$N$ unknown contribution;
    \item one function, $f_2$, is leading small-$N$ unknown contribution, which correspond to the highest power unknown for the pole at $N=1$;
    \item the remaining functions, $f_3$ and $f_4$, are chosen from a set of functions 
    describing sub-leading unknown terms both for the small-$N$ and large-$N$ limit.
\end{enumerate}
This way we generate a large set of independent candidates, around 70 for each anomalous dimension.
By taking the spread of the solutions we obtain as an estimate of the IHOUs
which can be added as a separate contribution to the theory covariance matrix.
The best result is always taken to be the average on all the possible variations.

Second, we apply a post-fit selection criteria to reduce the number of
candidates ($\approx 20$ for each function) selecting the most representative elements and discarding outlier
solutions. In particular, we adopt the following criteria:
\begin{itemize}
    \item Among the functions selected in step 3  above, we select candidates
    containing at least one of the leading sub-leading small-$N$ ($N=0,1$)
    or large-$N$ unknown contributions, such that the spread of the reduced ensemble is
    not smaller than the full one.
    \item By evaluating at the $x$-space line integral, we discard outliers
    that can be generated by large numerical cancellations.
\end{itemize}
Table~\ref{tab:1} illustrates the considered 
input functions\footnote{See \url{https://eko.readthedocs.io/en/latest/theory/N3LO_ad.html} for the full set.} for the reduced candidate sets for $\gamma_{gg}^{(3)}$.

\begin{table}[H]
    \centering
    \begin{tabular}{c|c}
        $f_1(N)$ & $\frac{S_1(N)}{N} \overset{\ N \to \infty}{\approx} \frac{\ln{N}}{N}$ \\
        \midrule
        $f_2(N)$ & $\frac{S_2(N-2)}{N} \overset{\ N \to 1}{\approx} \frac{1}{(N-1)^2}$ \\
        \midrule
        $f_3(N)$ & $\frac{1}{N-1},\ \frac{1}{N}$ \\
        \midrule
        $f_4(N)$ & $\frac{1}{N-1},\ \frac{1}{N^4},\ \frac{1}{N^3},\ \frac{1}{N^2},\ \frac{1}{N},\ \frac{1}{(N+1)^3},\ \frac{1}{(N+1)^2},\ \frac{1}{N+1},\ \frac{1}{N+2},$ \\
                 & $\mathcal{M}[(1-x)\ln(1-x)],\ \frac{S_1(N)}{N^2}$
    \end{tabular}
    \caption{Functions used in $\gamma_{gg}^{(3)}(x)$ to parametrize the difference
    between known functional dependence and the Mellin moments. \label{tab:1}}
\end{table}

\begin{figure}[h]
    \centering
    \includegraphics[width=0.48\textwidth]{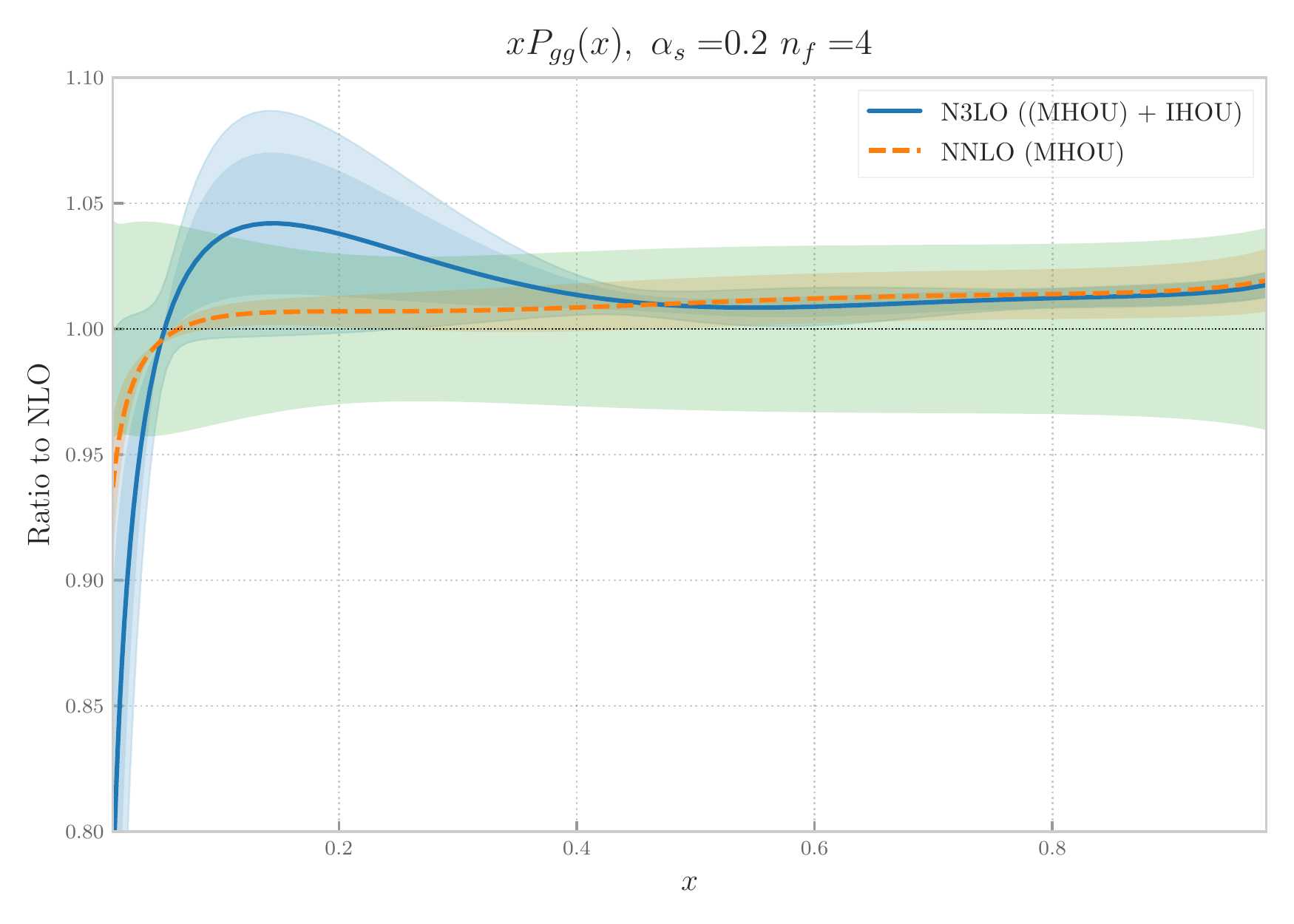}
    \includegraphics[width=0.48\textwidth]{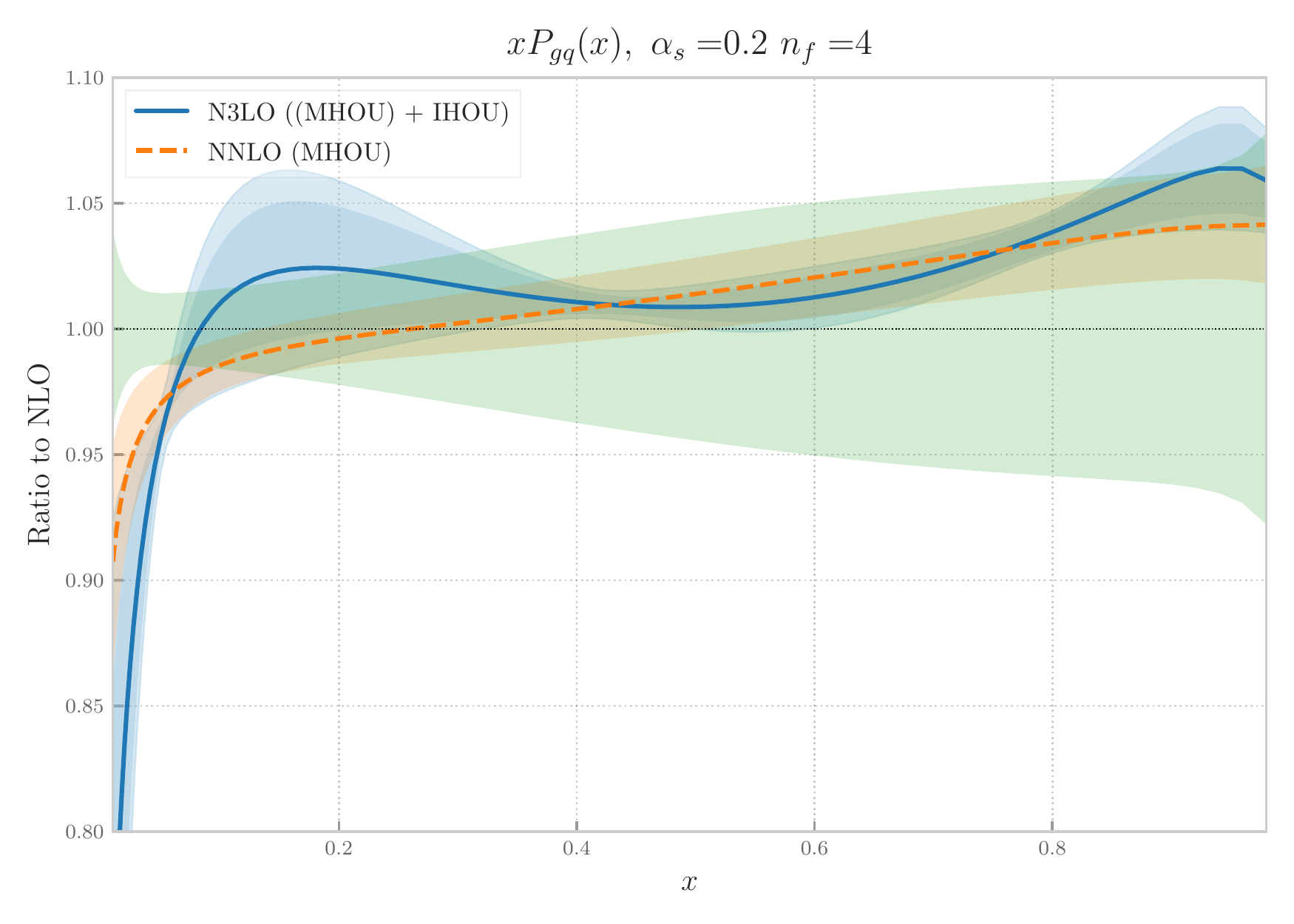} \\
    \includegraphics[width=0.48\textwidth]{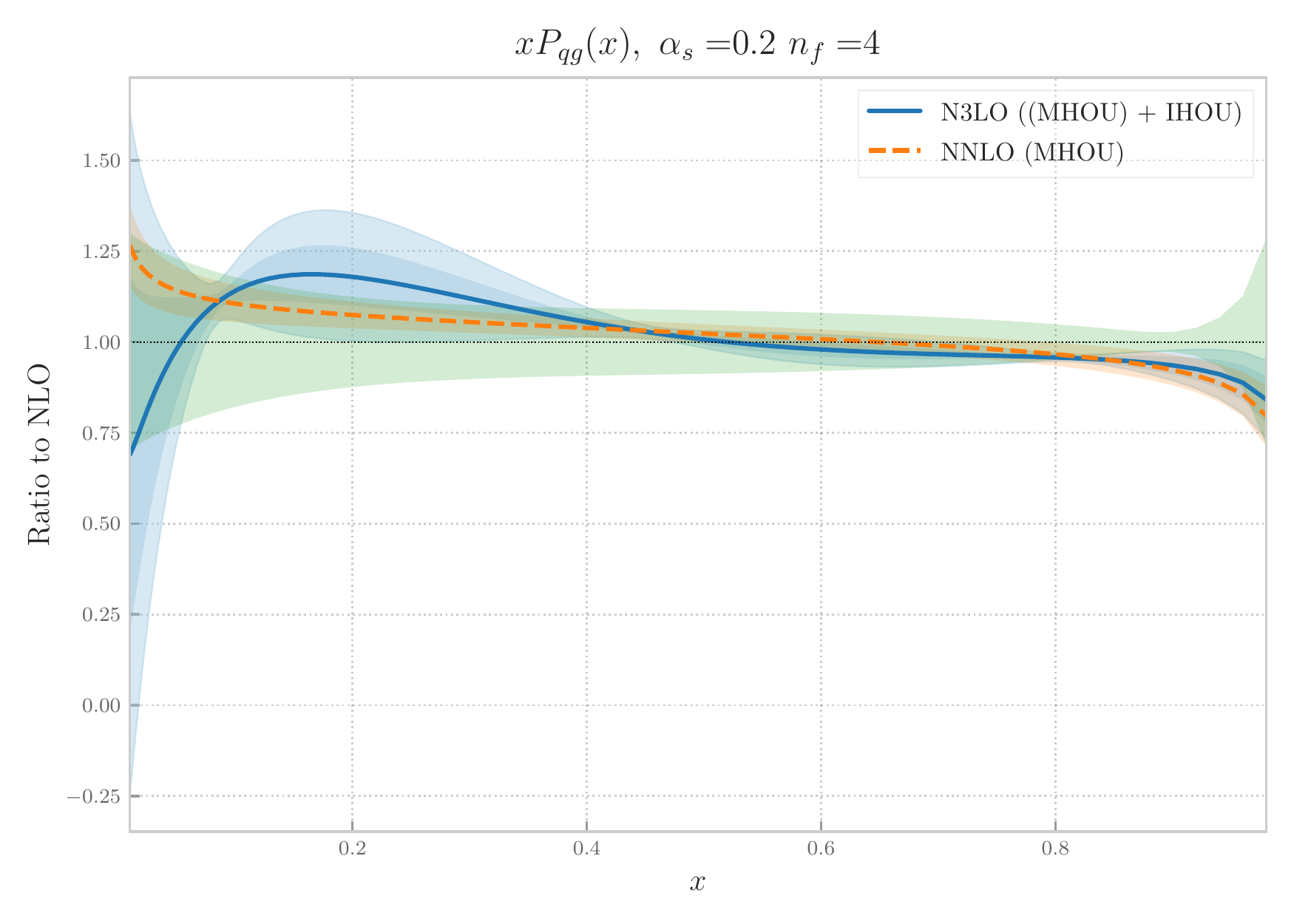}
    \includegraphics[width=0.48\textwidth]{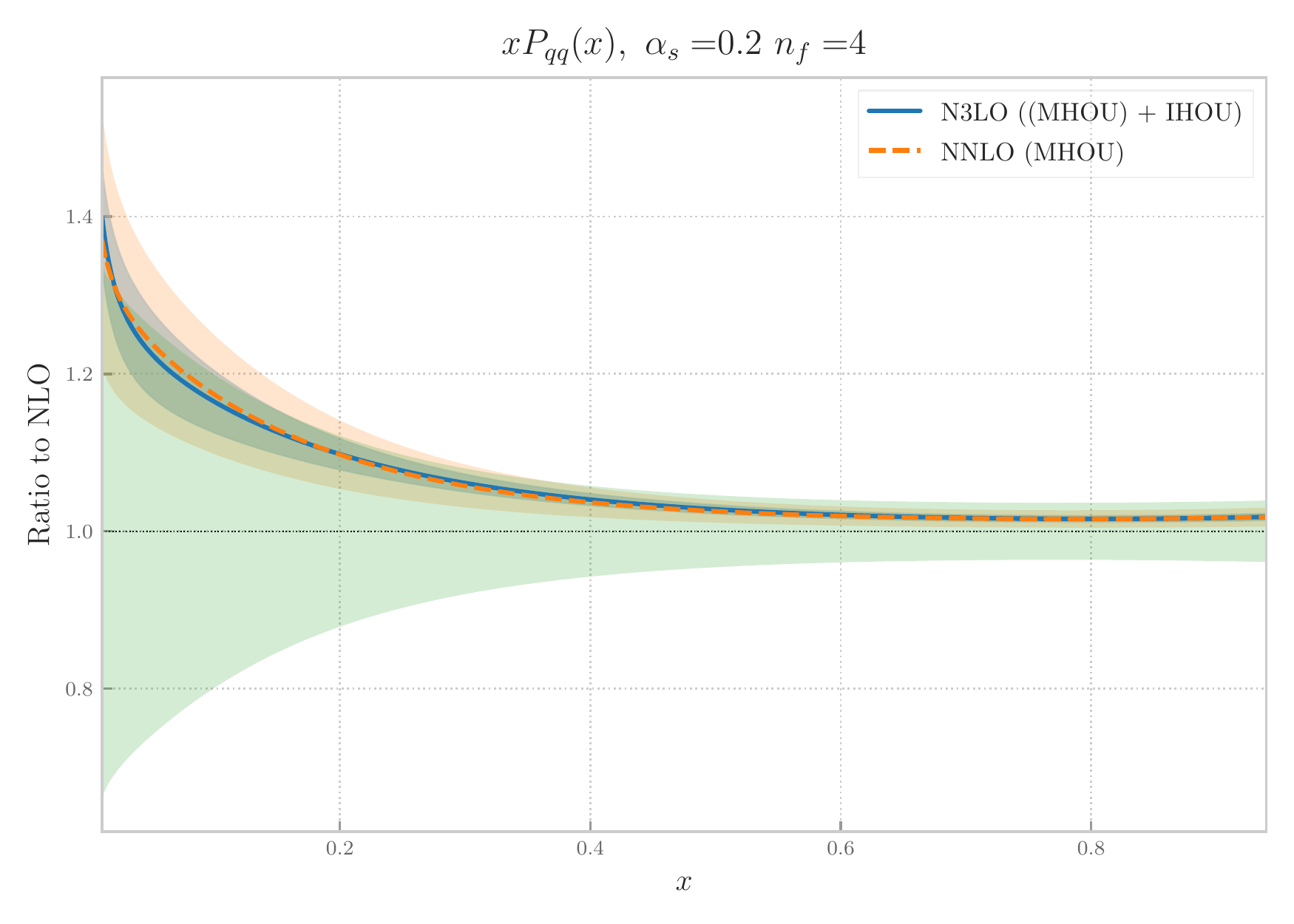}
    \caption{The singlet splitting functions at large-$x$ at different perturbative orders
    as a ratio to the NLO result.
    For the aN$^3$LO results, the dark error bands correspond to MHOU, while the lighter to IHOU.
    For NLO and NNLO, the error bands corresponds to MHOU.}
    \label{fig:mhou_linear}
\end{figure}

\begin{figure}[h]
    \centering
    \includegraphics[width=0.48\textwidth]{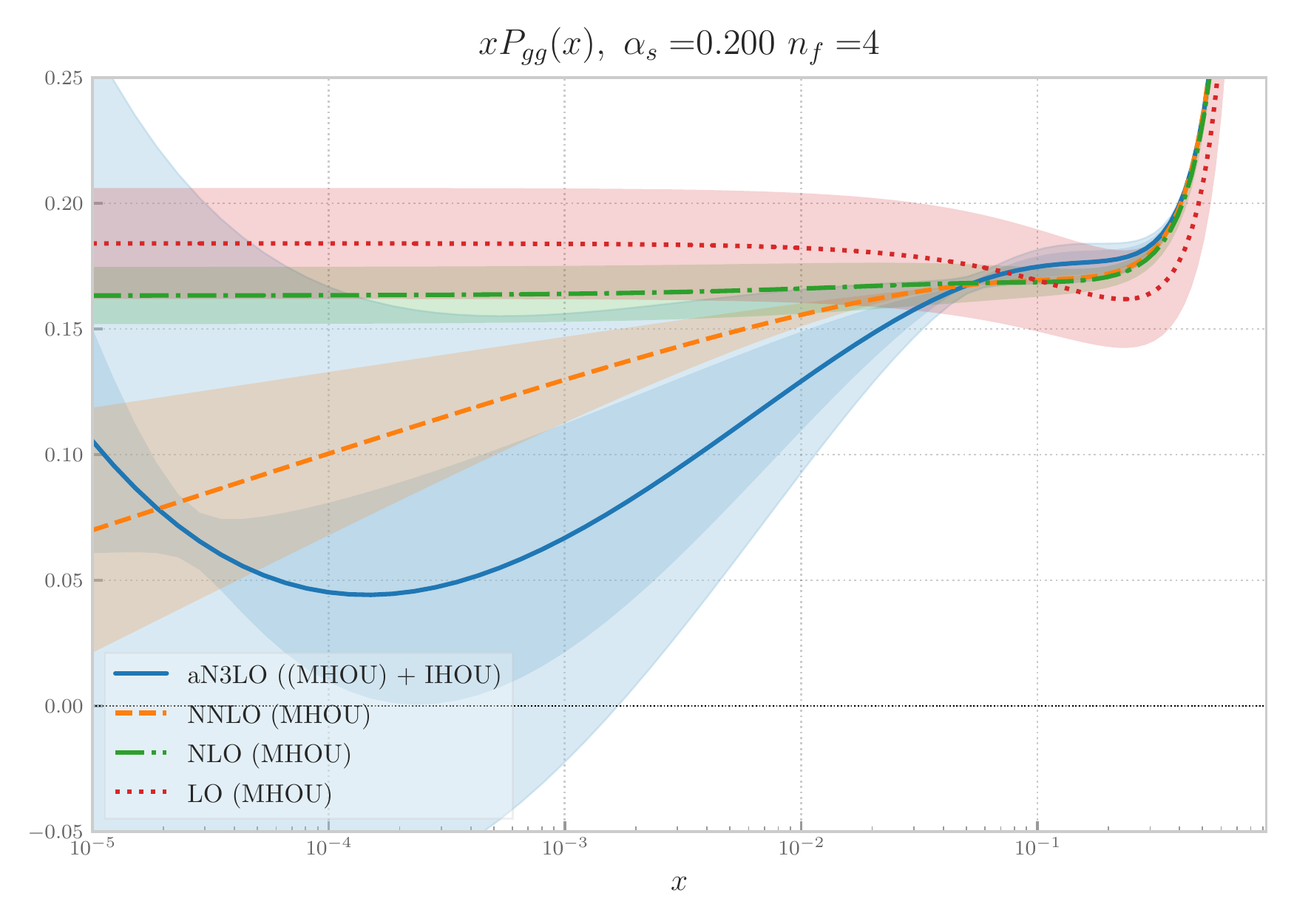}
    \includegraphics[width=0.48\textwidth]{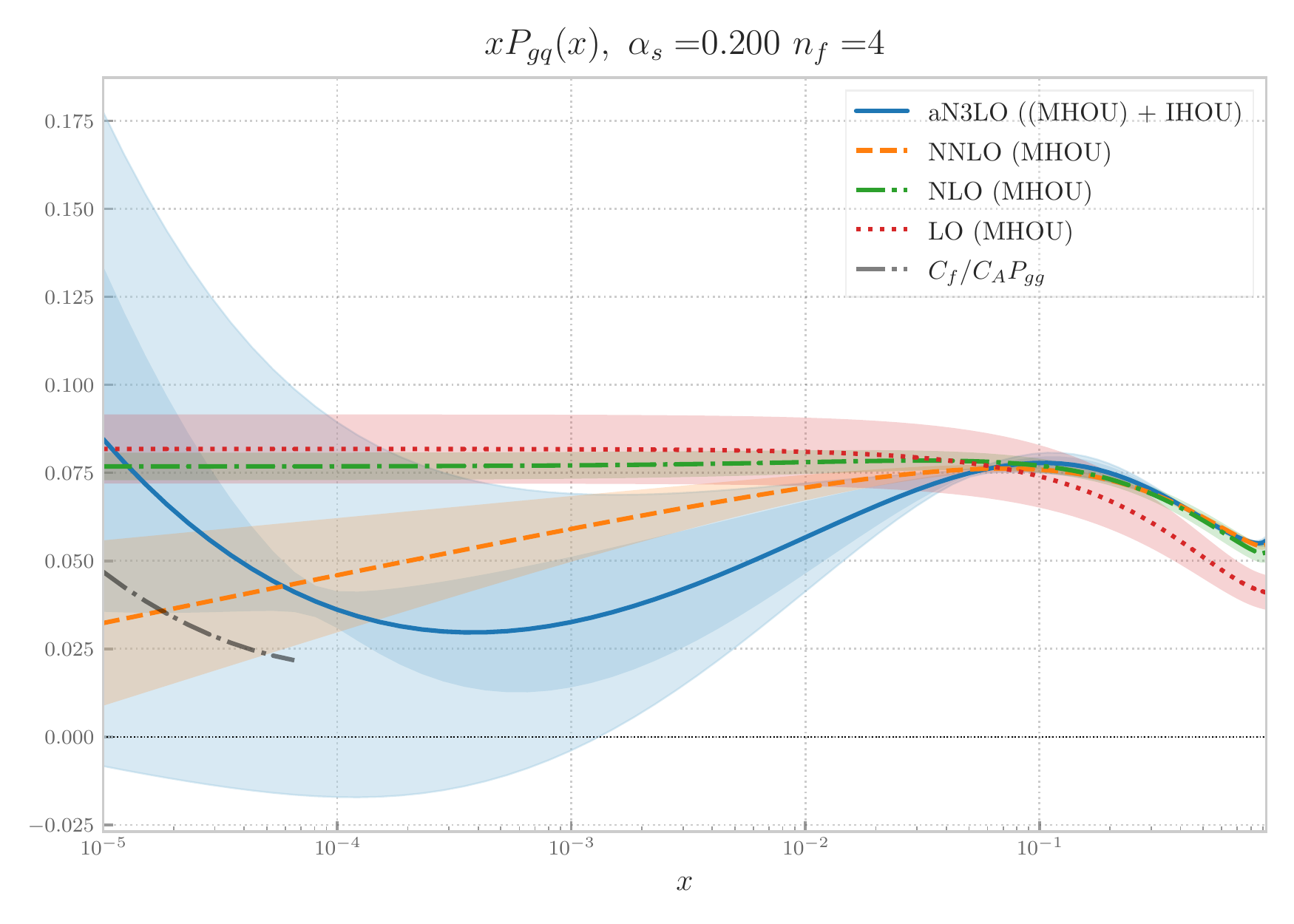} \\
    \includegraphics[width=0.48\textwidth]{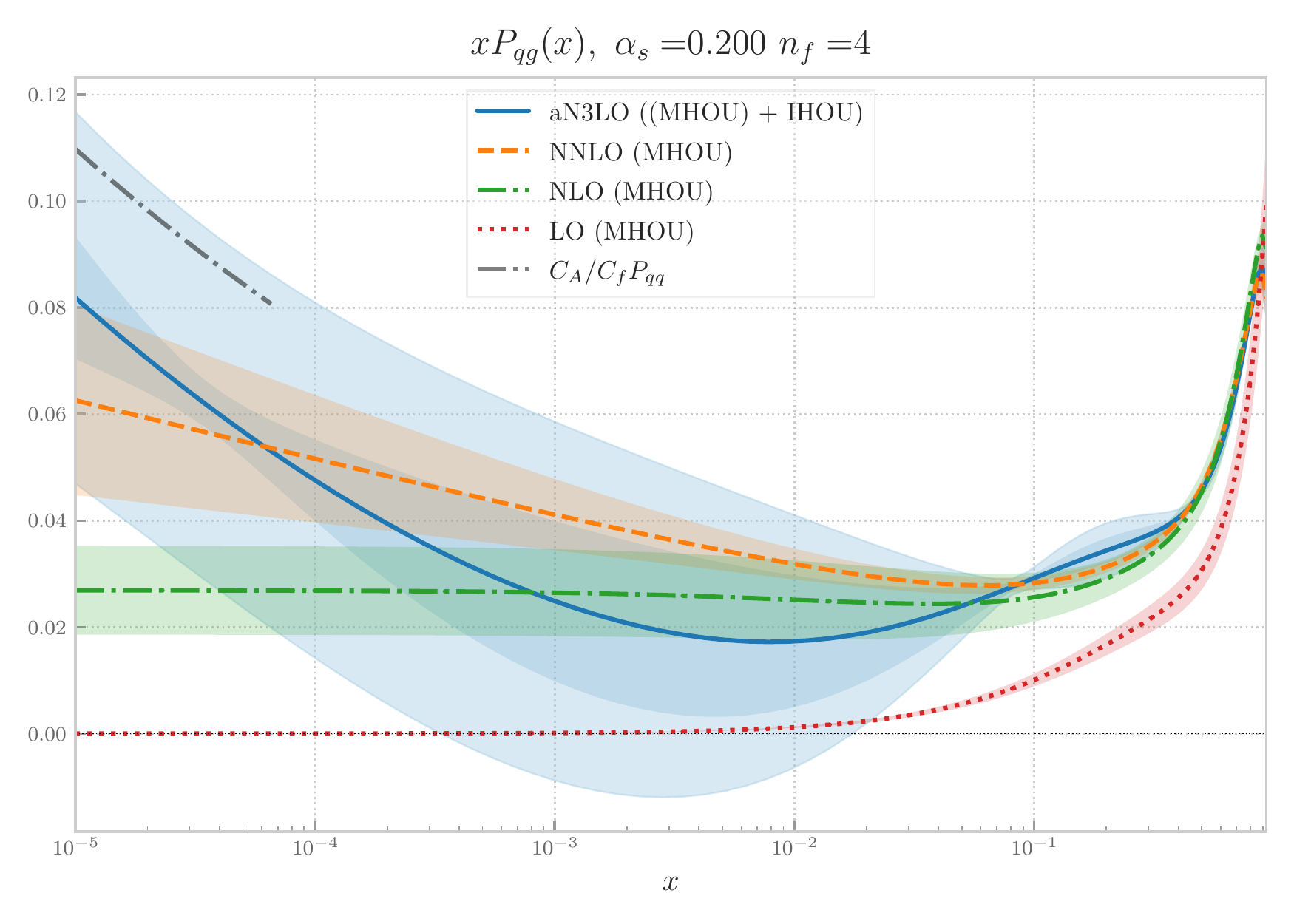}
    \includegraphics[width=0.48\textwidth]{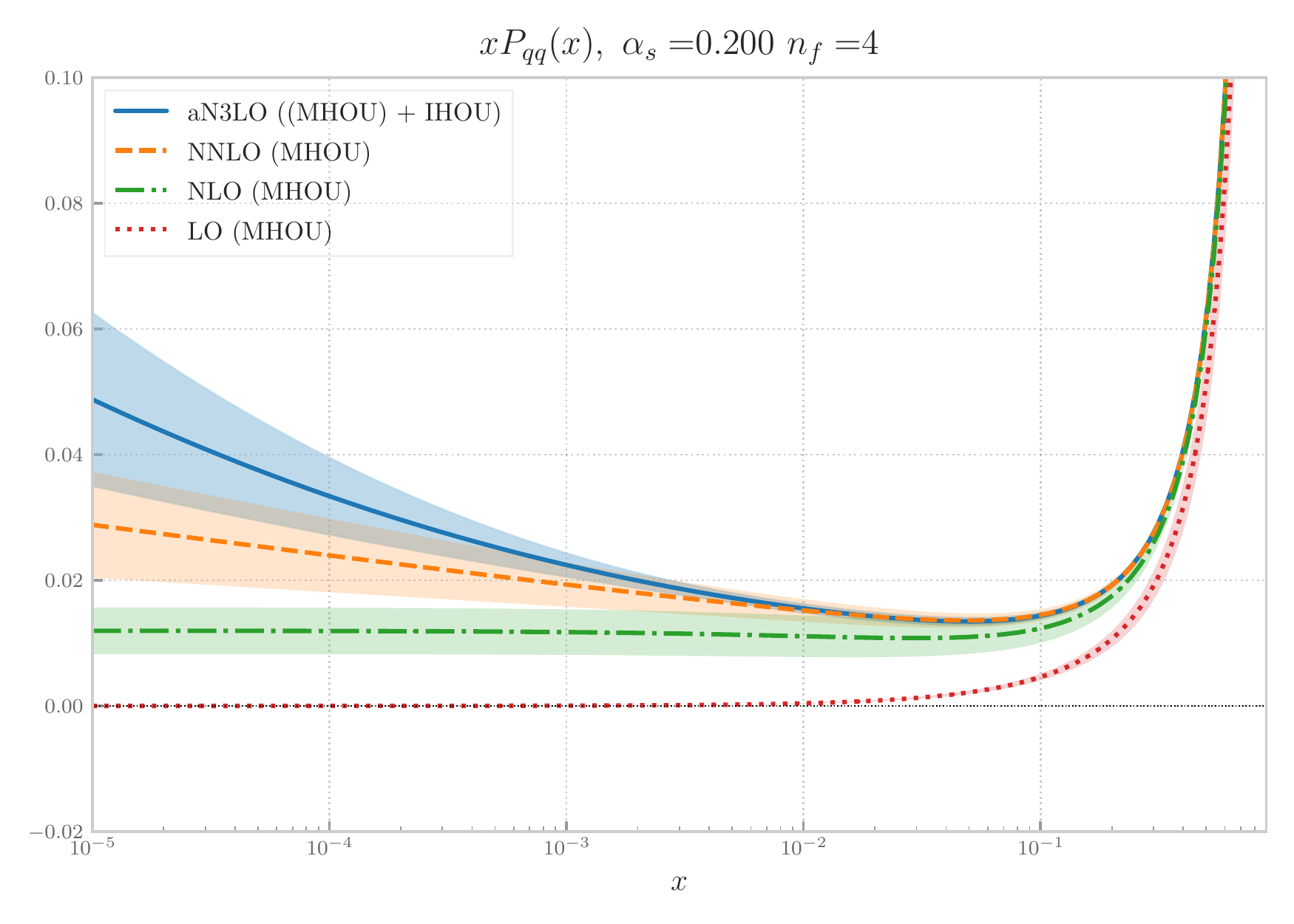}
    \caption{Same as Fig.~\ref{fig:mhou_linear}
 in a logarithmic scale, without taking
 the ratio, and adding the LO splitting functions.
    For the off-diagonal terms, the dashed line indicates the LL scaling.} 
    \label{fig:mhou_log}
\end{figure}

\paragraph{Results.}
In \cref{fig:mhou_linear,fig:mhou_log} we show the
resulting aN$^3$LO splitting functions,
both in linear and logarithmic scales respectively.
The uncertainties bands include MHOU up to NNLO and both MHOU (dark blue) 
and IHOU (light blue) for the approximate N$^3$LO.
MHOUs are obtained by varying $\rho \in [0.5,2]$ in \cref{eq:MHOU}.

In \cref{fig:mhou_linear} we observe the stable perturbative expansions for $x \ge 0.1$, 
with overlapping uncertainties at each order.
In \cref{fig:mhou_log} in the small-$x$ region this is not the case as large logarithms arise.
On the other hand, MHOUs, despite being unable to estimate the raise of new divergences,
are capturing the correct trend also in this region, with NNLO uncertainty always larger than NLO.
We remark also how the logarithmic series is alternating in sign with the NLL contributions
cancelling the LL one in intermediate $x$ regions, as clearly visible for $P_{gg}$ in \cref{fig:mhou_log}.
We note how the uncertainty band on $P_{qq}$ is fully dominated by the MHOU.
This confirms how the knowledge of 8 ($P_{NS,+}$) or 10 ($P_{qq,PS}$) Mellin moments is
sufficient to determine the N$^3$LO result with a sufficient accuracy for any phenomenological
study, despite not enough to reconstruct the full analytical structure.

The aN$^3$LO splitting functions and anomalous dimensions
displayed in \cref{fig:mhou_linear,fig:mhou_log}
are available in the open-source python package \texttt{EKO}\footnote{\url{https://github.com/NNPDF/eko}} and enable the DGLAP evolution
of the PDFs at aN$^3$LO.
They are being used to produce a variant of the NNPDF4.0 global
analysis based on aN$^3$LO QCD calculations which will be released
in the coming months.
%

\bibliography{main}

\providecommand{\href}[2]{#2}\begingroup\raggedright\begin{thebibliography}{10}

\bibitem{Amoroso:2022eow}
S.~Amoroso et~al., {\it {Snowmass 2021 Whitepaper: Proton Structure at the
  Precision Frontier}},  {\em Acta Phys. Polon. B} {\bf 53} (2022), no.~12 1,
  [\href{http://arxiv.org/abs/2203.13923}{{\tt arXiv:2203.13923}}].

\bibitem{Gao:2017yyd}
J.~Gao, L.~Harland-Lang, and J.~Rojo, {\it {The Structure of the Proton in the
  LHC Precision Era}},  {\em Phys. Rept.} {\bf 742} (2018) 1--121,
  [\href{http://arxiv.org/abs/1709.04922}{{\tt arXiv:1709.04922}}].

\bibitem{Caola:2022ayt}
F.~Caola, W.~Chen, C.~Duhr, X.~Liu, B.~Mistlberger, F.~Petriello, G.~Vita, and
  S.~Weinzierl, {\it {The Path forward to N$^3$LO}},  in {\em {2022 Snowmass
  Summer Study}}, 3, 2022.
\newblock \href{http://arxiv.org/abs/2203.06730}{{\tt arXiv:2203.06730}}.

\bibitem{McGowan:2022nag}
J.~McGowan, T.~Cridge, L.~A. Harland-Lang, and R.~S. Thorne, {\it {Approximate
  N$^{3}$LO parton distribution functions with theoretical uncertainties:
  MSHT20aN$^3$LO PDFs}},  {\em Eur. Phys. J. C} {\bf 83} (2023), no.~3 185,
  [\href{http://arxiv.org/abs/2207.04739}{{\tt arXiv:2207.04739}}]. [Erratum:
  Eur.Phys.J.C 83, 302 (2023)].

\bibitem{Moch:2004pa}
S.~Moch, J.~A.~M. Vermaseren, and A.~Vogt, {\it {The Three loop splitting
  functions in QCD: The Nonsinglet case}},  {\em Nucl. Phys.} {\bf B688} (2004)
  101--134, [\href{http://arxiv.org/abs/hep-ph/0403192}{{\tt hep-ph/0403192}}].

\bibitem{Vogt:2004mw}
A.~Vogt, S.~Moch, and J.~A.~M. Vermaseren, {\it {The Three-loop splitting
  functions in QCD: The Singlet case}},  {\em Nucl. Phys.} {\bf B691} (2004)
  129--181, [\href{http://arxiv.org/abs/hep-ph/0404111}{{\tt hep-ph/0404111}}].

\bibitem{Blumlein:2021enk}
J.~Bl\"umlein, P.~Marquard, C.~Schneider, and K.~Sch\"onwald, {\it {The
  three-loop unpolarized and polarized non-singlet anomalous dimensions from
  off shell operator matrix elements}},  {\em Nucl. Phys. B} {\bf 971} (2021)
  115542, [\href{http://arxiv.org/abs/2107.06267}{{\tt arXiv:2107.06267}}].

\bibitem{Davies:2016jie}
J.~Davies, A.~Vogt, B.~Ruijl, T.~Ueda, and J.~A.~M. Vermaseren, {\it
  {Large-$n_f$ contributions to the four-loop splitting functions in QCD}},
  {\em Nucl. Phys. B} {\bf 915} (2017) 335--362,
  [\href{http://arxiv.org/abs/1610.07477}{{\tt arXiv:1610.07477}}].

\bibitem{Moch:2017uml}
S.~Moch, B.~Ruijl, T.~Ueda, J.~A.~M. Vermaseren, and A.~Vogt, {\it {Four-Loop
  Non-Singlet Splitting Functions in the Planar Limit and Beyond}},  {\em JHEP}
  {\bf 10} (2017) 041, [\href{http://arxiv.org/abs/1707.08315}{{\tt
  arXiv:1707.08315}}].

\bibitem{Bonvini:2018xvt}
M.~Bonvini and S.~Marzani, {\it {Four-loop splitting functions at small $x$}},
  {\em JHEP} {\bf 06} (2018) 145, [\href{http://arxiv.org/abs/1805.06460}{{\tt
  arXiv:1805.06460}}].

\bibitem{Davies:2022ofz}
J.~Davies, C.~H. Kom, S.~Moch, and A.~Vogt, {\it {Resummation of small-x double
  logarithms in QCD: inclusive deep-inelastic scattering}},
  \href{http://arxiv.org/abs/2202.10362}{{\tt arXiv:2202.10362}}.

\bibitem{Moch:2021qrk}
S.~Moch, B.~Ruijl, T.~Ueda, J.~A.~M. Vermaseren, and A.~Vogt, {\it {Low moments
  of the four-loop splitting functions in QCD}},  {\em Phys. Lett. B} {\bf 825}
  (2022) 136853, [\href{http://arxiv.org/abs/2111.15561}{{\tt
  arXiv:2111.15561}}].

\bibitem{Falcioni:2023luc}
G.~Falcioni, F.~Herzog, S.~Moch, and A.~Vogt, {\it {Four-loop splitting
  functions in QCD -- The quark-quark case}},
  \href{http://arxiv.org/abs/2302.07593}{{\tt arXiv:2302.07593}}.

\bibitem{AbdulKhalek:2019ihb}
{\bf NNPDF} Collaboration, R.~Abdul~Khalek et~al., {\it {Parton Distributions
  with Theory Uncertainties: General Formalism and First Phenomenological
  Studies}},  {\em Eur. Phys. J. C} {\bf 79} (2019), no.~11 931,
  [\href{http://arxiv.org/abs/1906.10698}{{\tt arXiv:1906.10698}}].

\bibitem{AbdulKhalek:2019bux}
{\bf NNPDF} Collaboration, R.~Abdul~Khalek et~al., {\it {A First Determination
  of Parton Distributions with Theoretical Uncertainties}},
  \href{http://arxiv.org/abs/1905.04311}{{\tt arXiv:1905.04311}}.

\bibitem{Candido:2022tld}
A.~Candido, F.~Hekhorn, and G.~Magni, {\it {EKO: evolution kernel operators}},
  {\em Eur. Phys. J. C} {\bf 82} (2022), no.~10 976,
  [\href{http://arxiv.org/abs/2202.02338}{{\tt arXiv:2202.02338}}].

\bibitem{Barontini:2023vmr}
A.~Barontini, A.~Candido, J.~M. Cruz-Martinez, F.~Hekhorn, and C.~Schwan, {\it
  {Pineline: Industrialization of High-Energy Theory Predictions}},
  \href{http://arxiv.org/abs/2302.12124}{{\tt arXiv:2302.12124}}.

\bibitem{NNPDF:2021uiq}
{\bf NNPDF} Collaboration, R.~D. Ball et~al., {\it {An open-source machine
  learning framework for global analyses of parton distributions}},  {\em Eur.
  Phys. J. C} {\bf 81} (2021), no.~10 958,
  [\href{http://arxiv.org/abs/2109.02671}{{\tt arXiv:2109.02671}}].

\bibitem{Ball:2021leu}
{\bf NNPDF} Collaboration, R.~D. Ball et~al., {\it {The path to proton
  structure at 1\% accuracy}},  {\em Eur. Phys. J. C} {\bf 82} (2022), no.~5
  428, [\href{http://arxiv.org/abs/2109.02653}{{\tt arXiv:2109.02653}}].

\bibitem{Thorne:1999rb}
R.~S. Thorne, {\it {NLO BFKL equation, running coupling and renormalization
  scales}},  {\em Phys. Rev. D} {\bf 60} (1999) 054031,
  [\href{http://arxiv.org/abs/hep-ph/9901331}{{\tt hep-ph/9901331}}].

\bibitem{Ball:2005mj}
R.~D. Ball and S.~Forte, {\it {All order running coupling BFKL evolution from
  GLAP (and vice-versa)}},  {\em Nucl. Phys. B} {\bf 742} (2006) 158--175,
  [\href{http://arxiv.org/abs/hep-ph/0601049}{{\tt hep-ph/0601049}}].

\bibitem{Ball:2017otu}
R.~D. Ball, V.~Bertone, M.~Bonvini, S.~Marzani, J.~Rojo, and L.~Rottoli, {\it
  {Parton distributions with small-x resummation: evidence for BFKL dynamics in
  HERA data}},  {\em Eur. Phys. J. C} {\bf 78} (2018), no.~4 321,
  [\href{http://arxiv.org/abs/1710.05935}{{\tt arXiv:1710.05935}}].

\bibitem{Albino:2000cp}
S.~Albino and R.~D. Ball, {\it {Soft resummation of quark anomalous dimensions
  and coefficient functions in MS-bar factorization}},  {\em Phys. Lett. B}
  {\bf 513} (2001) 93--102, [\href{http://arxiv.org/abs/hep-ph/0011133}{{\tt
  hep-ph/0011133}}].

\bibitem{Duhr:2022cob}
C.~Duhr, B.~Mistlberger, and G.~Vita, {\it {Soft integrals and soft anomalous
  dimensions at N$^{3}$LO and beyond}},  {\em JHEP} {\bf 09} (2022) 155,
  [\href{http://arxiv.org/abs/2205.04493}{{\tt arXiv:2205.04493}}].

\bibitem{Soar:2009yh}
G.~Soar, S.~Moch, J.~A.~M. Vermaseren, and A.~Vogt, {\it {On Higgs-exchange
  DIS, physical evolution kernels and fourth-order splitting functions at large
  x}},  {\em Nucl. Phys. B} {\bf 832} (2010) 152--227,
  [\href{http://arxiv.org/abs/0912.0369}{{\tt arXiv:0912.0369}}].

\bibitem{Vogt:2004ns}
A.~Vogt, {\it {Efficient evolution of unpolarized and polarized parton
  distributions with QCD-PEGASUS}},  {\em Comput. Phys. Commun.} {\bf 170}
  (2005) 65--92, [\href{http://arxiv.org/abs/hep-ph/0408244}{{\tt
  hep-ph/0408244}}].

\end{thebibliography}\endgroup
 
\end{document}